\documentclass{article}

\usepackage{microtype}
\usepackage{graphicx}
\usepackage{subfigure}
\usepackage{booktabs} 
\usepackage{makecell}
\usepackage{hyperref}
\usepackage{enumitem}
\setlist[itemize]{itemsep=0pt, topsep=0pt}

\usepackage[accepted]{icml2025}

\usepackage{amsmath}
\usepackage{amssymb}
\usepackage{mathtools}
\usepackage{amsthm}

\usepackage{amsmath,amsfonts,bm}

\def\eqref#1{equation~\ref{#1}}

\def\1{\bm{1}}

 \def\d{\mathrm{d}}

\def\vzero{{\bm{0}}}

\def\vepsilon{{\bm{\epsilon}}}

\def\vs{{\bm{s}}}

\def\vx{{\bm{x}}}

\def\mI{{\bm{I}}}

\DeclareMathAlphabet{\mathsfit}{\encodingdefault}{\sfdefault}{m}{sl}
\SetMathAlphabet{\mathsfit}{bold}{\encodingdefault}{\sfdefault}{bx}{n}

\newcommand{\E}{\mathbb{E}}

\usepackage[capitalize,noabbrev]{cleveref}

\theoremstyle{plain}
\newtheorem{theorem}{Theorem}[section]
\newtheorem{proposition}[theorem]{Proposition}

\theoremstyle{definition}

\theoremstyle{remark}

\usepackage[textsize=tiny]{todonotes}
\usepackage{multicol}
\usepackage{multirow}

\icmltitlerunning{Revisiting Sampling Strategies for Molecular Generation}

\begin{document}

\twocolumn[
\icmltitle{Revisiting Sampling Strategies for Molecular Generation}




\begin{icmlauthorlist}
\icmlauthor{Yuyan Ni}{scha,schb,scht}
\icmlauthor{Shikun Feng}{scht}
\icmlauthor{Wei-Ying Ma}{scht}
\icmlauthor{Zhi-Ming Ma}{scha}
\icmlauthor{Yanyan Lan}{scht,sche}
\end{icmlauthorlist}

\icmlaffiliation{scht}{Institute for AI Industry Research (AIR), Tsinghua University (Work was done during Yuyan's internship at AIR.)}
\icmlaffiliation{scha}{Academy of Mathematics and Systems Science, Chinese Academy of Sciences }
\icmlaffiliation{schb}{University of Chinese Academy of Sciences}
\icmlaffiliation{sche}{Beijing Academy of Artificial Intelligence}

\icmlcorrespondingauthor{Yanyan Lan}{lanyanyan@air.tsinghua.edu.cn}

\icmlkeywords{Machine Learning, ICML}

\vskip 0.3in
]



\printAffiliationsAndNotice{}  

\begin{abstract}
Sampling strategies in diffusion models are critical to molecular generation yet remain relatively underexplored. In this work, we investigate a broad spectrum of sampling methods beyond conventional defaults and reveal that sampling choice substantially affects molecular generation performance. In particular, we identify a maximally stochastic sampling (StoMax), a simple yet underexplored strategy, as consistently outperforming default sampling methods for generative models DDPM and BFN. Our findings highlight the pivotal role of sampling design and suggest promising directions for advancing molecular generation through principled and more expressive sampling approaches.
\end{abstract}

\section{Introduction}
Molecular generation has emerged as a crucial task in AI-driven drug and material discovery, enabling the rapid exploration of chemical space for novel and functional compounds. Among the many approaches to generative modeling, diffusion models have recently achieved state-of-the-art performance in 3D molecular generation tasks due to their superior capability in precisely modeling atomic positions. 

While extensive research has focused on improving model architectures and training objectives, a key component that remains relatively under-explored is the sampling strategy, the procedure by which new molecules are generated from the learned diffusion models.
In most existing approaches, including Equivariant Diffusion Models (EDM) \cite{hoogeboom2022equivariant} and Geometric Bayesian Flow Networks (GeoBFN) \cite{song2023geobfn}, the sampling process is inherently tied to the model’s design. Specifically, these methods adopt the default sampling method of Denoising Diffusion Probabilistic Models (DDPM) \cite{ho2020denoising} and Bayesian Flow Networks (BFN) \cite{graves2023bayesian}, which correspond to first-order discretizations of the reverse-time stochastic differential equations (SDEs) of the diffusion process. 

However, the default sampling strategies commonly used in diffusion models are not the only theoretically valid choices. In fact, a broader family of sampling methods exists, each defined by distinct assumptions about temporal dependencies in the generative process. These variations can substantially affect the quality of generated samples, and hold untapped potential for improving molecular generation performance.

To further comprehend this design space, we first identify two representative cases that conclude widely adopted strategies:
(1) a Markov forward process, which underlies DDPM and BFN samplers, and
(2) a Deterministic reverse process, corresponding to DDIM \cite{songjiam2020denoising} and ODE-based sampling \cite{song2021scorebased}.
Beyond these well-studied methods, we introduce a third, intuitive alternative:
(3) a conditionally independent reverse process, in which the noisy sample at each reverse timestep is conditionally independent of the previous timestep given the initial data. This method induces maximal stochasticity in the family of sampling methods, we termed maximally stochastic sampling (StoMax). 

We systematically evaluate StoMax, a previously underexplored sampling alternative, across multiple molecular generative models, including UniGEM(EDM), UniGEM(BFN) \cite{feng2024unigem}, EDM, and GeoBFN. Empirically, StoMax consistently outperforms the native samplers of these models on both the QM9 and GEOM-Drugs datasets. Notably, StoMax brings substantial improvements under the DDPM noise schedule, fully leveraging the capacity of pretrained generative models. 
This enables UniGEM(EDM) to achieve state-of-the-art performance in molecular generation.
Interestingly, DDIM and StoMax represent the two extremes of the sampling family in terms of stochasticity, with DDPM and BFN default sampling falling in between. 
To probe the potential of this design space, we interpolate between the three sampling strategies and find that StoMax yields the best overall sample quality, with a minor trade-off in diversity. 

These empirical observations motivate us to explore more diverse and expressive sampling strategies beyond conventional choices, aiming to further enhance the quality of molecular generation. The results also motivate the development of theoretical frameworks that can explain these empirical gains and guide the design of optimal samplers.

\section{Revisiting Sampling Methods in Diffusion Models}
\subsection{General Reverse SDE}
Following the notation introduced in \citet{ni2025sldm}, we describe the noise corruption process in the diffusion model as:
\begin{equation}\label{eq:xt def}\small
\vx_t=\mu_t\vx_0+\sigma_t\vepsilon, \quad \vepsilon\sim \mathcal{N}(\vzero,\mI_N),
\end{equation}
where $\vx_0, \vx_t \in \mathbb{R}^N$ denote the clean and corrupted data at time $t$, respectively. The functions $\mu_t$ and $\sigma_t$ define the noise schedule, with $t \in [0, T]$. This general framework can include DDPM and BFN. For DDPM, it satisfies the variance preserving (VP) assumption $\mu_t^2+\sigma_t^2=1$, while for BFN on continuous data, we have $\sigma_t=\sqrt{\mu_t(1-\mu_t)}$.

This process corresponds to a linear stochastic differential equation (SDE) \citep{ni2025sldm}:
\begin{equation}\label{eq:sde}\small
\mathrm{d}\vx_t = \frac{\dot{\mu}_t}{\mu_t} \vx_t \mathrm{d}t + \sqrt{2\sigma_t\dot{\sigma}_t - 2\sigma_t^2 \frac{\dot{\mu}_t}{\mu_t}}  \mathrm{d}w_t,
\end{equation}
where $\mathrm{d}w_t$ denotes the standard Wiener process. 

 Interestingly, it is possible to design \textbf{a family of reverse processes} (Prop. 4.1 in \citet{xue2024sa}) that share the same marginal probability distributions as \eqref{eq:sde} :
\begin{equation}\label{eq:reverse-sde}\small
\begin{aligned}    
    d\vx_t=&\left[\frac{\dot{\mu}_t}{\mu_t}\vx_t-\frac{1+\beta(t)}{2}g^2(t)\nabla_x\log p_t(x)\right]\d t\\&
    +\sqrt{\beta(t)}g_t\d w_t,
\end{aligned}
\end{equation}
where $g_t=\sqrt{2}\sqrt{\sigma_t\dot{\sigma}_t-\sigma_t^2\frac{\dot{\mu}_t}{\mu_t}}$, $\beta(t)$ is any non-negative bounded function. 
As proved in Appendix E of \citet{song2021scorebased} and Proposition 4.2 in \citet{xue2024unifyingbfn} respectively, the sampling processes of both DDPM and BFN can be interpreted as first-order discretizations of the reverse-time SDE in \eqref{eq:reverse-sde} with $\beta=1$.


\subsection{Discretized Sampling Strategy with Varying Correlation Hypotheses}
To derive concrete sampling strategies, we begin by discretizing the reverse-time SDE in \eqref{eq:reverse-sde}. Although different discretization schemes may lead to variations in the resulting sampling formula, these differences become negligible when the step size is sufficiently small. As this work focuses on improving sampling quality in the regime of a large number of discretization steps, we adopt a simple first-order discretization:
\begin{equation}\small \label{eq:discrete reverse sde}
    \begin{aligned}
        \vx_{t-\Delta t}&=\frac{\mu_{t-\Delta t}}{\mu_t}\vx_t+ \frac{(1+\beta(t))g_t^2\Delta t}{2}\nabla_x\log p_t(\vx_t)\\& +\sqrt{\beta(t)g_t^2\Delta t}\vepsilon,
    \end{aligned}
    \end{equation}
where $g_t^2\Delta t\approx 2\sigma_t^2\frac{\mu_{t-\Delta t}}{\mu_t}-2\sigma_t\sigma_{t-\Delta t}$, with derivation given in Appendix \ref{app：discretize}.
In this formulation, the next state is sampled from a Gaussian distribution conditioned on the current state. The optimal mean of this distribution, i.e. the conditional expectation $\E[\vx_{t-\Delta t}|\vx_t]$ has an analytic form involving the score function. This expectation depends on the form of the conditional distribution $p(\vx_{t-\Delta t}|\vx_t,\vx_0)$ which captures the correlation between different time steps. According to \cite{songjiam2020denoising}, a family of such conditional distributions parameterized by $\lambda_t$ all satisfy \eqref{eq:xt def}:
\begin{equation}\small\label{eq:ddim}
\begin{aligned}
    p_\lambda(x_{t-\Delta t}|x_t,x_0)=N(\mu_{t-\Delta t}x_0+\gamma_t\frac{x_t-\mu_t x_0}{\sigma_t},\lambda_t^2I), 
\end{aligned}
\end{equation} where $\gamma_t^2+\lambda_t^2=\sigma_{t-\Delta t}^2$.   
Based on this, the conditional expectation is given by: 
\begin{equation}\small\label{eq:mean}
\begin{aligned}
\E[\vx_{t-\Delta t}|\vx_t]=\frac{\mu_{t-\Delta t}}{\mu_t}\vx_t +(\frac{\mu_{t-\Delta t}}{\mu_t}\sigma_t^2-\gamma_t\sigma_t)\nabla_x\log p_t(\vx_t),
\end{aligned}
\end{equation}
with derivation given in Appendix \ref{app:Analytic Expectation}.
By aligning the mean in ~\eqref{eq:discrete reverse sde} with this conditional expectation, we derive the corresponding variance as $2\sigma_t(\sigma_{t-\Delta t}-\gamma_t)$. Substituting this into the original expression, the discretized reverse SDE becomes:
\begin{equation}\small \label{eq:exact discrete reverse sde}
    \begin{aligned}
        \vx_{t-\Delta t}&=\frac{\mu_{t-\Delta t}}{\mu_t}\vx_t +(\frac{\mu_{t-\Delta t}}{\mu_t}\sigma_t-\gamma_t)\sigma_t\nabla_x\log p_t(\vx_t) \\&+\sqrt{2\sigma_t(\sigma_{t-\Delta t}-\gamma_t)}\vepsilon. 
    \end{aligned}
    \end{equation}

\subsection{Canonical Sampling Methods and Beyond}
In this subsection, we examine the landscape of feasible sampling strategies, each reflecting a different temporal correlation assumption. First of all, we identify two representative cases that together cover the most widely used sampling strategies in existing diffusion models: 



\begin{itemize}
    \item Markov forward process:
When the forward diffusion process satisfies the Markov property, i.e., $p(\vx_t|\vx_{t-\Delta t},\vx_0) = p(\vx_t|\vx_{t-\Delta t})$, we prove in the Proposition \ref{prop: markov} and Proposition \ref{prop: bfn markov} that this corresponds to the choice 
$\gamma_t=\frac{\mu_t\sigma_{t-\Delta t}^2}{\mu_{t-\Delta t}\sigma_t}$. Note that this result does not depend on the VP assumption and both DDPM and BFN sampling belong to this category.
    \item Deterministic reverse process:
When the reverse process becomes deterministic, i.e. $p(\vx_{t-\Delta t}|\vx_t,\vx_0)$ collapses to a point mass, this corresponds to setting $\gamma_t=\sigma_{t-\Delta t}$, as in DDIM sampling.
\end{itemize} 


These two methods span a spectrum of correlation structures from fully stochastic to fully deterministic. However, it remains an open and compelling question whether alternative, potentially superior sampling strategies exist within this family.
We now consider a less-explored but intuitively natural alternative that lies at the opposite end of the deterministic method:
\begin{itemize}
    \item When the reverse process is conditional independent, that is $p(x_{n-1}|x_n,x_0)=p(x_{n-1}|x_0)$. This corresponds to setting $\gamma_t=0$, which maximizes the variance in both \eqref{eq:ddim} and \eqref{eq:exact discrete reverse sde}. We refer to this setting as maximally stochastic sampling (StoMax).
\end{itemize} 
Please note that the categorization here refers to how the mean is chosen during the iterative process, while allowing for multiple theoretically sound options for the variance. By default, we use the variance from the SDE sampling formulation in \eqref{eq:exact discrete reverse sde}. However, alternative choices, such as those used in DDIM or Analytic-DPM \cite{bao2022analytic}, are also feasible.
In the deterministic case, these methods all yield zero variance. In the Markovian case, the performance of different variance choices tends to converge as the number of discrete steps increases \cite{bao2022analytic}. In the StoMax setting, the variance can be interpreted as a form of temperature control \cite{ni2025sldm}, offering a trade-off between sample diversity and fidelity. 


\section{Experimental Results}
In this section, we compare the performance of different sampling strategies on molecular generation tasks. 

\begin{figure*}[t]
	\centering
    \includegraphics[width=1\linewidth]{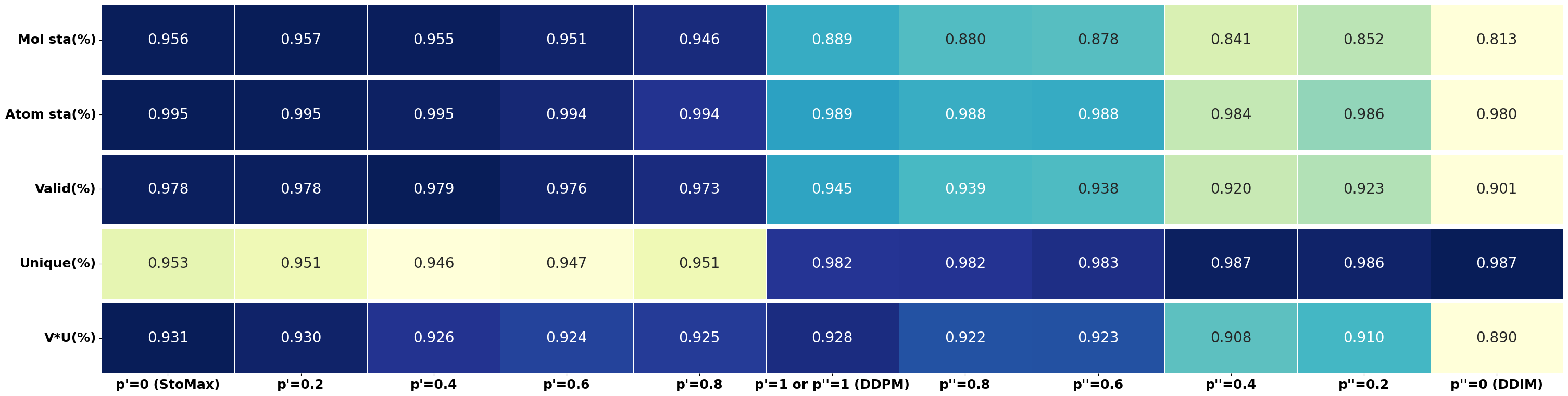}
    \vskip-1pt
    \caption{Unconditional molecular generation results on QM9.
We evaluate interpolation-based sampling strategies across StoMax, DDPM, and DDIM using the UniGem (EDM) model. Higher values indicate better performance across all metrics.}
    \label{fig:interpolate}
\end{figure*}
\subsection{Settings}

\textbf{Datasets} We conduct generation experiments on two commonly used datasets. The QM9 dataset \cite{ruddigkeit2012enumeration, ramakrishnan2014quantum} is a database containing approximately 134,000 organic small molecules, composed of C, H, O, N, and F atoms. Each molecule contains up to 9 heavy atoms. We adopt the same data split strategy as the baseline methods, with 100k molecules in the training set, 18k in the validation set, and 13k in the test set.
The other dataset is GEOM-Drugs \cite{axelrod2022geom}, which consists of drug-like molecules. It contains 430,000 molecules, with an average of 44 atoms per molecule and up to 181 atoms in the largest molecule, making it more challenging than QM9. The dataset is split in the same way as in previous works: randomly divided into training, validation, and test sets with a ratio of 8:1:1.

\textbf{Baselines} Our baselines include the classic 3D molecular generation algorithm EDM \cite{hoogeboom2022equivariant}, which performs joint diffusion over discrete atom types and continuous molecular coordinates. GeoBFN \cite{song2023geobfn} improves the scheduling of discrete atom types and continuous coordinates using Bayesian Flow Networks. UniGEM \cite{feng2024unigem} treats molecular generation as a two-stage process: a nucleation phase for generating molecular scaffolds and a growth phase for completing the molecule. In UniGEM, atom type prediction is decoupled and performed only during the growth phase. For coordinate generation, UniGEM can adopt either EDM or GeoBFN, and we refer to these variants as UniGEM(EDM) and UniGEM(GeoBFN), respectively.

\begin{table}[t]
\caption{Unconditional molecular generation results on QM9. For all diffusion-based models, the sampling steps are $1000$. Metrics are calculated with 10000 samples generated from each model. Higher values indicate better performance. *: The GeoBFN model is re-trained and evaluated by ourselves, as the original paper did not release the pretrained model.}
\label{exp:main_gen}
\vspace{5pt}
\setlength{\tabcolsep}{1pt}
\centering
\scalebox{0.93}{
\begin{tabular}{l|l|cccc}
 \toprule
\multicolumn{1}{l|}{Models} &Sampling& Atom sta(\%) & Mol sta(\%) & Valid(\%)   & V*U(\%)  \\
\hline
\multicolumn{1}{l|}{Data}  &-     & 99.0 &95.2& 97.7  & 97.7       \\
\hline
\multirow{3}{*}{EDM} &Default       & 98.7& 82.0        & 91.9         & 90.7  \\
                            &StoMax & \textbf{98.9}&\textbf{87.9}&\textbf{94.5}&\textbf{92.1}\\
                            \hline
\multirowcell{3}[4pt][l]{UniGEM\\(EDM)}&Default & 99.0 & 89.8 & 95.0 & 93.2\\
                            &StoMax & \textbf{99.6}&\textbf{96.1}&\textbf{98.1}&\textbf{93.7}\\
                            \hline
\multirow{3}{*}{GeoBFN} &Default*      & 99.3 & 93.0 & 96.5&         \textbf{92.7} \\
                         &StoMax & \textbf{99.3} & \textbf{94.2} & \textbf{96.9} &91.9\\
                            \hline
\multirowcell{3}[4pt][l]{UniGEM\\(BFN)}& Default &  99.3  & 93.7 & 97.3 & \textbf{93.0}  \\
                            &StoMax & \textbf{99.5} & \textbf{95.7} & \textbf{97.8} & 91.3\\

\bottomrule

\end{tabular}}
\end{table}

\begin{table}[t]
\centering
\caption{Comparison between the StoMax strategy and the default unconditional sampling for EDM and UniGEM on the GEOM-Drugs dataset. 10,000 molecules were sampled using 1,000 diffusion steps.}
\label{exp:main_geomdrug}
\vspace{5pt}
\begin{tabular}{l|l|c|c}
\toprule
\multicolumn{1}{l|}{Models} & Sampling & Atom sta(\%) & Valid(\%) \\
\hline
\multicolumn{1}{l|}{Data}  & -         & 86.5     & 99.9      \\
\hline

\multirow{3}{*}{EDM}         & Default   & 81.3     & 92.6      \\
                             & StoMax   & \textbf{86.2} & \textbf{99.7} \\
\hline
\multirow{3}{*}{UniGEM(EDM)} & Default   & 85.1     & 98.4      \\
                             & StoMax   & \textbf{89.5} & \textbf{99.9} \\

\bottomrule
\end{tabular}
\end{table}

\textbf{Metric} We follow the evaluation protocol of prior works \cite{hoogeboom2022equivariant}, generating 10,000 molecules to assess atom stability, molecule stability, validity, and valid \& unique (V*U). Consistent with EDM’s strategy, covalent bonds are inferred from inter-atomic distances. Atom stability is the portion of atoms with correct valence; molecule stability is the portion of molecules whose atoms all satisfy valence rules. Validity measures the proportion of generated generated 3D structures convertible to valid SMILES via RDKit. The V*U metric calculates the proportion of unique samples among all valid molecules.

\subsection{Molecular Generation Evaluated on QM9 Dataset}
We applied the StoMax strategy to EDM, GeoBFN, and two variants of UniGEM (UniGEM(EDM) and UniGEM(BFN)) and evaluated their unconditional generation performance on the QM9 dataset. The results are summarized in Table \ref{exp:main_gen}. Here default sampling for EDM and GeoBFN refers to the sampling method of DDPM and BFN, respectively. 
StoMax consistently improves quality-related metrics, atom and molecule stability, and validity across all models, demonstrating its general effectiveness in enhancing sample quality. The gains are especially pronounced for models based on EDM.
A slight drop in diversity is observed, particularly for GeoBFN and UniGEM(GeoBFN), suggesting a trade-off between stability and uniqueness.

\subsection{Molecular Generation Evaluated on GEOM-Drugs}
We evaluate StoMax on the larger and more structurally diverse GEOM-Drugs dataset. As shown in Table~\ref{exp:main_geomdrug}, we apply StoMax to EDM and UniGEM(EDM) and compare the results of unconditional generation with their respective default samplers. StoMax yields notable gains in sample quality. Atom stability improves by around 5 points for EDM and over 4 points for UniGEM. In addition, the validity metric was improved to near-perfect levels (close to 1.0) in both models, clearly highlighting the advantage of StoMax in generating high-quality molecular samples.

\subsection{Interpolation}

To investigate whether better sampling methods exist within the explored design space, we follow \cite{songjiam2020denoising} and conduct interpolation experiments between different sampling strategies. Notably, the deterministic and StoMax approaches represent two extremes of the sampling family in terms of stochasticity, with the sampling method based on a Markovian forward process lying between them.
We thus interpolate between StoMax and DDPM using $\gamma_t=p'\gamma_t^{\text{DDPM}} +(1-p')\gamma_t^{\text{StoMax}} =  p'\frac{\mu_t\sigma_{t-\Delta t}^2}{\mu_{t-\Delta t}\sigma_t} $, and between DDPM and DDIM via $\gamma_t=p''\gamma_t^{\text{DDPM}} +(1-p'')\gamma_t^{\text{DDIM}} =  p''\frac{\mu_t\sigma_{t-\Delta t}^2}{\mu_{t-\Delta t}\sigma_t}+(1-p'')\sigma_{t-\Delta t} $, where $p'$ and $p''$ are interpolation coefficients in $[0, 1]$. We evaluate the resulting sampling strategies using the UniGEM(EDM) model.

As shown in Figure~\ref{fig:interpolate}, increasing the level of stochasticity consistently enhances the stability and validity of generated molecules, though it slightly compromises uniqueness.
Notably, the StoMax strategy achieves the highest score on the U×V metric, demonstrating a favorable trade-off between diversity and generation quality. Among the sampling methods considered, StoMax emerges as the most balanced and effective approach.


\section{Discussion and Future Work}

Our findings raise compelling questions about the theoretical foundations of optimal sampling in diffusion-based generative models. While Appendix C in \cite{xue2024sa} suggest that the optimal sampling strategy that minimizes the ELBO corresponds to  $\beta(t)=1$ in \eqref{eq:reverse-sde}, which aligns with the default samplers used in DDPM and BFN, our empirical results demonstrate that, at least for molecular generation, alternative strategies such as StoMax can yield superior performance.

Despite the strong empirical performance of StoMax on molecular tasks, its rigorous theoretical explanation is still lacking. Interestingly, viewing the continuous formulation in \eqref{eq:reverse-sde}, the StoMax strategy, characterized by maximal sampling variance, can be seen as approximating Langevin dynamics by taking the limit $\beta(t) \to \infty$, although our discrete implementation still differs from the commonly used Langevin dynamics used in generative models \cite{song2019scoremat,saremi2019wjs}. This observation raises the intriguing possibility that Langevin-like samplers may be inherently better suited for molecule generation. However, this remains a hypothesis that requires further systematic investigation.

Moreover, our experiments reveal a clear trade-off between diversity and validity across different sampling strategies. Developing a unified theoretical framework to characterize how different variance schedules affect this trade-off remains an open and promising direction for future work.


\section*{Impact Statement}
This paper presents work whose goal is to advance the field of 
Machine Learning. There are many potential societal consequences 
of our work, none which we feel must be specifically highlighted here.

\nocite{langley00}

\bibliography{example_paper}

\begin{thebibliography}{16}
\providecommand{\natexlab}[1]{#1}
\providecommand{\url}[1]{\texttt{#1}}
\expandafter\ifx\csname urlstyle\endcsname\relax
  \providecommand{\doi}[1]{doi: #1}\else
  \providecommand{\doi}{doi: \begingroup \urlstyle{rm}\Url}\fi

\bibitem[Axelrod \& Gomez-Bombarelli(2022)Axelrod and Gomez-Bombarelli]{axelrod2022geom}
Axelrod, S. and Gomez-Bombarelli, R.
\newblock Geom, energy-annotated molecular conformations for property prediction and molecular generation.
\newblock \emph{Scientific Data}, 9\penalty0 (1):\penalty0 185, 2022.

\bibitem[Bao et~al.(2022)Bao, Li, Zhu, and Zhang]{bao2022analytic}
Bao, F., Li, C., Zhu, J., and Zhang, B.
\newblock Analytic-dpm: an analytic estimate of the optimal reverse variance in diffusion probabilistic models.
\newblock \emph{International conference on learning representations}, 2022.

\bibitem[Feng et~al.(2025)Feng, Ni, Lu, Ma, Ma, and Lan]{feng2024unigem}
Feng, S., Ni, Y., Lu, Y., Ma, Z.-M., Ma, W.-Y., and Lan, Y.
\newblock Unigem: A unified approach to generation and property prediction for molecules.
\newblock \emph{International conference on learning representations}, 2025.

\bibitem[Graves et~al.(2023)Graves, Srivastava, Atkinson, and Gomez]{graves2023bayesian}
Graves, A., Srivastava, R.~K., Atkinson, T., and Gomez, F.
\newblock Bayesian flow networks.
\newblock \emph{arXiv preprint arXiv:2308.07037}, 2023.

\bibitem[Ho et~al.(2020)Ho, Jain, and Abbeel]{ho2020denoising}
Ho, J., Jain, A., and Abbeel, P.
\newblock Denoising diffusion probabilistic models.
\newblock \emph{Advances in neural information processing systems}, 33:\penalty0 6840--6851, 2020.

\bibitem[Hoogeboom et~al.(2022)Hoogeboom, Satorras, Vignac, and Welling]{hoogeboom2022equivariant}
Hoogeboom, E., Satorras, V.~G., Vignac, C., and Welling, M.
\newblock Equivariant diffusion for molecule generation in 3d.
\newblock In \emph{International conference on machine learning}, pp.\  8867--8887. PMLR, 2022.

\bibitem[Ni et~al.(2025)Ni, Feng, Chi, Zheng, Gao, Ma, Ma, and Lan]{ni2025sldm}
Ni, Y., Feng, S., Chi, H., Zheng, B., Gao, H.-a., Ma, W.-Y., Ma, Z.-M., and Lan, Y.
\newblock Straight-line diffusion model for efficient 3d molecular generation.
\newblock \emph{arXiv preprint arXiv:2503.02918}, 2025.

\bibitem[Ramakrishnan et~al.(2014)Ramakrishnan, Dral, Rupp, and Von~Lilienfeld]{ramakrishnan2014quantum}
Ramakrishnan, R., Dral, P.~O., Rupp, M., and Von~Lilienfeld, O.~A.
\newblock Quantum chemistry structures and properties of 134 kilo molecules.
\newblock \emph{Scientific data}, 1\penalty0 (1):\penalty0 1--7, 2014.

\bibitem[Ruddigkeit et~al.(2012)Ruddigkeit, Van~Deursen, Blum, and Reymond]{ruddigkeit2012enumeration}
Ruddigkeit, L., Van~Deursen, R., Blum, L.~C., and Reymond, J.-L.
\newblock Enumeration of 166 billion organic small molecules in the chemical universe database gdb-17.
\newblock \emph{Journal of chemical information and modeling}, 52\penalty0 (11):\penalty0 2864--2875, 2012.

\bibitem[Saremi \& Hyv{\"a}rinen(2019)Saremi and Hyv{\"a}rinen]{saremi2019wjs}
Saremi, S. and Hyv{\"a}rinen, A.
\newblock Neural empirical bayes.
\newblock \emph{Journal of Machine Learning Research}, 20\penalty0 (181):\penalty0 1--23, 2019.

\bibitem[Song et~al.(2021{\natexlab{a}})Song, Meng, and Ermon]{songjiam2020denoising}
Song, J., Meng, C., and Ermon, S.
\newblock Denoising diffusion implicit models.
\newblock 2021{\natexlab{a}}.
\newblock URL \url{https://openreview.net/forum?id=St1giarCHLP}.

\bibitem[Song \& Ermon(2019)Song and Ermon]{song2019scoremat}
Song, Y. and Ermon, S.
\newblock Generative modeling by estimating gradients of the data distribution.
\newblock \emph{Advances in neural information processing systems}, 32, 2019.

\bibitem[Song et~al.(2021{\natexlab{b}})Song, Sohl-Dickstein, Kingma, Kumar, Ermon, and Poole]{song2021scorebased}
Song, Y., Sohl-Dickstein, J., Kingma, D.~P., Kumar, A., Ermon, S., and Poole, B.
\newblock Score-based generative modeling through stochastic differential equations.
\newblock In \emph{International Conference on Learning Representations}, 2021{\natexlab{b}}.
\newblock URL \url{https://openreview.net/forum?id=PxTIG12RRHS}.

\bibitem[Song et~al.(2023)Song, Gong, Zhou, Zheng, Liu, and Ma]{song2023geobfn}
Song, Y., Gong, J., Zhou, H., Zheng, M., Liu, J., and Ma, W.-Y.
\newblock Unified generative modeling of 3d molecules with bayesian flow networks.
\newblock In \emph{The Twelfth International Conference on Learning Representations}, 2023.

\bibitem[Xue et~al.(2024{\natexlab{a}})Xue, Zhou, Nie, Min, Zhang, ZHOU, and Li]{xue2024unifyingbfn}
Xue, K., Zhou, Y., Nie, S., Min, X., Zhang, X., ZHOU, J., and Li, C.
\newblock Unifying bayesian flow networks and diffusion models through stochastic differential equations.
\newblock In \emph{Forty-first International Conference on Machine Learning}, 2024{\natexlab{a}}.
\newblock URL \url{https://openreview.net/forum?id=1jHiq640y1}.

\bibitem[Xue et~al.(2024{\natexlab{b}})Xue, Yi, Luo, Zhang, Sun, Li, and Ma]{xue2024sa}
Xue, S., Yi, M., Luo, W., Zhang, S., Sun, J., Li, Z., and Ma, Z.-M.
\newblock Sa-solver: Stochastic adams solver for fast sampling of diffusion models.
\newblock \emph{Advances in Neural Information Processing Systems}, 36, 2024{\natexlab{b}}.

\end{thebibliography}
\bibliographystyle{icml2025}

\newpage
\appendix
\onecolumn
\section{Complementary Proofs}

\subsection{Discretizing the Reverse-Time SDE}\label{app：discretize}
To derive the discrete-time update rule for sampling, we begin by discretizing the reverse-time SDE in ~\eqref{eq:reverse-sde} using the Euler–Maruyama method. The resulting update step takes the following form:
    \begin{equation}\small \label{eqapp:family update}
    \begin{aligned}
        \vx_{t-\Delta t}&=\vx_t- \left[\frac{\dot{\mu}_t}{\mu_t}\vx_t- \frac{(1+\beta(t))g_t^2}{2}\nabla_x\log p_t(\vx_t) 
 \right]\Delta t+\sqrt{\beta(t)g_t^2\Delta t}\vepsilon,
    \end{aligned}
    \end{equation}
    where $\vepsilon\sim \mathcal{N}(\vzero,\mI_N)$ denotes Gaussian noise.

    To simplify the expression, we expand $\mu_{t - \Delta t}$ and $\sigma_{t - \Delta t}$ using a first-order Taylor approximation:
    \begin{equation}
        \mu_{t-\Delta t}=\mu_t-\dot{\mu}_t\Delta t+O((\Delta t)^2), \sigma_{t-\Delta t}=\sigma_t-\dot{\sigma}_t\Delta t+O((\Delta t)^2)
    \end{equation}    
    Recall that $g_t^2 = 2 \left( \sigma_t \dot{\sigma}_t - \sigma_t^2 \frac{\dot{\mu}_t}{\mu_t} \right)$, applying the above Taylor expansions, we approximate: 
    \begin{equation}\label{eq:g_delta_t}
    g_t^2\Delta t= 2\sigma_t\dot{\sigma}_t\Delta t-2\sigma_t^2\frac{\dot{\mu}_t}{\mu_t}\Delta t \approx 2\sigma_t^2\frac{\mu_{t-\Delta t}}{\mu_t}-2\sigma_t\sigma_{t-\Delta t}.
    \end{equation}

    Substituting this into~\eqref{eqapp:family update}, and collecting terms, we obtain the final discrete-time update:
    \begin{equation} \label{eqapp:discrete reverse sde}
    \begin{aligned}
        \vx_{t-\Delta t}&=\frac{\mu_{t-\Delta t}}{\mu_t}\vx_t+ \frac{(1+\beta(t))g_t^2\Delta t}{2}\nabla_x\log p_t(\vx_t) +\sqrt{\beta(t)g_t^2\Delta t}\vepsilon,
    \end{aligned}
    \end{equation}
    where $g_t^2\Delta t\approx 2\sigma_t^2\frac{\mu_{t-\Delta t}}{\mu_t}-2\sigma_t\sigma_{t-\Delta t}$.
    This formulation offers a general discrete-time sampling framework encompassing a broad family of strategies, each determined by specific choices of $\beta(t)$

   \subsection{Analytic Expectation of the Reverse Probability}\label{app:Analytic Expectation}
We begin by connecting the score function to the conditional expectation of the original data:
\begin{equation}\label{eq: score and expectation}
    \begin{aligned}
        \nabla_{\vx_t}\log p_t(\vx_t)&=\frac{\int f(\vx_0) \nabla_{\vx_t} p(\vx_t|\vx_0)  d\vx_0}{p(\vx_t)}  
        = \frac{\int f(\vx_0) p(\vx_t|\vx_0) (-\frac{\vx_t-\mu_t \vx_0}{\sigma_t^2})  d\vx_0}{p(\vx_t)}  
        = -\frac{\vx_t}{\sigma_t^2}+\frac{\mu_t }{\sigma_t^2}\E[\vx_0|\vx_t],
    \end{aligned}
\end{equation}
which directly yields the Tweedie formula:
     \begin{equation}
    \E[\vx_0|\vx_t]=  \frac{1}{\mu_t}\left(\vx_t+\sigma_t^2\nabla_{\vx_t}\log p_t(\vx_t)\right)      
    \end{equation}
    
    To compute the expectation of the reverse sample $\vx_{t - \Delta t}$ given $\vx_t$, we write:
    \begin{equation}\label{eq:conditional exp}
    \begin{aligned}        
        \E[\vx_{t-\Delta t}|\vx_t]&=\int \vx_{t-\Delta t}p(\vx_{t-\Delta t}|\vx_t)\d \vx_{t-\Delta t}
        =\int \vx_{t-\Delta t}\int p(\vx_{t-\Delta t}|\vx_t,\vx_0)p(\vx_0|\vx_t)\d \vx_0\d \vx_{t-\Delta t}\\
        &=\int \left(\int \vx_{t-\Delta t}p(\vx_{t-\Delta t}|\vx_t,\vx_0) \d \vx_{t-\Delta t}\right) p(\vx_0|\vx_t)\d \vx_0     
    \end{aligned}
    \end{equation}
    Using \eqref{eq:ddim}, we know that: $\E[\vx_{t-\Delta t}|\vx_t,\vx_0]=\mu_{t-\Delta t}x_0+\gamma_t\frac{x_t-\mu_t x_0}{\sigma_t}$, where $\gamma_t=\sqrt{\sigma_{t-\Delta t}^2-\lambda_t^2}$.
    
    Thus \eqref{eq:conditional exp} can be further reduced to
\begin{equation}
    \begin{aligned}     
    \E[\vx_{t-\Delta t}|\vx_t] = \mu_{t-\Delta t}\E[\vx_0|\vx_t]+\gamma_t\frac{x_t-\mu_t \E[\vx_0|\vx_t]}{\sigma_t}
    \end{aligned}
    \end{equation}
    
    Substituting $\E[\vx_0|\vx_t]$ from \eqref{eq: score and expectation}, we have:
    \begin{equation}
         \E_p[\vx_{t-\Delta t}|\vx_t] = 
        \vx_{t-\Delta t}=\frac{\mu_{t-\Delta t}}{\mu_t}\vx_t +(\frac{\mu_{t-\Delta t}}{\mu_t}\sigma_t-\gamma_t)\sigma_t\nabla_x\log p_t(\vx_t) 
    \end{equation}
    thus we proved  \eqref{eq:exact discrete reverse sde}.

\subsection{Reverse-Time SDE with Analytic Expectation}\label{app：align}
By aligning the mean in \eqref{eq:discrete reverse sde} with \eqref{eq:mean}, we obtain:
\begin{equation}
    \begin{aligned}
        \frac{(1+\beta(t))g_t^2\Delta t}{2}&=(\frac{\mu_{t-\Delta t}}{\mu_t}\sigma_t-\gamma_t)\sigma_t\\
        \beta(t)g_t^2\Delta t&=2(\frac{\mu_{t-\Delta t}}{\mu_t}\sigma_t-\gamma_t)\sigma_t-g_t^2\Delta t\\
        &\approx 2(\frac{\mu_{t-\Delta t}}{\mu_t}\sigma_t-\gamma_t)\sigma_t-(2\sigma_t^2\frac{\mu_{t-\Delta t}}{\mu_t}-2\sigma_t\sigma_{t-\Delta t})\\
        &=2\sigma_t(\sigma_{t-\Delta t}-\gamma_t)
    \end{aligned}
\end{equation}
The approximation in the third line follows from \eqref{eq:g_delta_t}. This result provides the sampling variance and confirms the expression in \eqref{eq:exact discrete reverse sde}.

\subsection{Sampling with Markov Forward Process Assumption}\label{app:Markov forward Sampling}
\begin{proposition} \label{prop: markov}
    
Assume the forward diffusion process satisfies the Markov property:
\begin{equation}
    p(\vx_t | \vx_{t-\Delta t}, \vx_0) = p(\vx_t |\vx_{t-\Delta t}).
\end{equation}  
We show that this leads to the variance parameter in \eqref{eq:ddim} becomes:
\begin{equation}\label{eq:markov lambda}
    \lambda_t = \frac{\sigma_{t-\Delta t}}{\sigma_t} \sqrt{ \sigma_t^2 - \frac{\mu_t^2}{\mu_{t-\Delta t}^2} \sigma_{t-\Delta t}^2 }.
\end{equation}
Then $\gamma_t=\sqrt{\sigma_{t-\Delta t}^2-\lambda_t^2}=\frac{\sigma_{t-\Delta t}^2\mu_t}{\sigma_t\mu_{t-\Delta t}}$.

\end{proposition}
   \begin{proof}
        Following the derivation approach in \eqref{eq:ddim}, the conditional distribution $p(\vx_t | \vx_{t-\Delta t}, \vx_0)$ under general variance $\tilde{\lambda}_t$ takes the form:
\begin{equation}\label{eq:ddim-reverse}
    p(\vx_t | \vx_{t-\Delta t}, \vx_0) = \mathcal{N} \left(
        \mu_t \vx_0 + \tilde{\gamma}_{t-\Delta t} \cdot \frac{\vx_{t-\Delta t} - \mu_{t-\Delta t} \vx_0}{\sigma_{t-\Delta t}},
        \tilde{\lambda}_{t-\Delta t}^2 \mathbf{I}
    \right),
\end{equation}
where $\tilde{\gamma}_{t-\Delta t} = \sqrt{\sigma_t^2 - \tilde{\lambda}_{t-\Delta t}^2}$.

Now, consider the identity:
\begin{equation}
    p(\vx_t \mid \vx_{t-\Delta t}, \vx_0) \, p(\vx_{t-\Delta t} \mid \vx_0) = p(\vx_{t-\Delta t} \mid \vx_t, \vx_0) \, p(\vx_t \mid \vx_0),
\end{equation}
and substitute the Gaussian expressions into both sides. By matching the exponents, we obtain the following equality:
\begin{align}
    &\frac{\left\| \vx_{t-\Delta t} - \mu_{t-\Delta t} \vx_0 - \gamma_t \cdot \frac{\vx_t - \mu_t \vx_0}{\sigma_t} \right\|^2}{2 \lambda_t^2}
    + \frac{\left\| \vx_t - \mu_t \vx_0 \right\|^2}{2 \sigma_t^2} 
    \\&= 
    \frac{\left\| \vx_t - \mu_t \vx_0 - \tilde{\gamma}_{t-\Delta t} \cdot \frac{\vx_{t-\Delta t} - \mu_{t-\Delta t} \vx_0}{\sigma_{t-\Delta t}} \right\|^2}{2 \tilde{\lambda}_{t-\Delta t}^2}
    + \frac{\left\| \vx_{t-\Delta t} - \mu_{t-\Delta t} \vx_0 \right\|^2}{2 \sigma_{t-\Delta t}^2}.
\end{align}

Treating $\vx_t - \mu_t \vx_0$ and $\vx_{t-\Delta t} - \mu_{t-\Delta t} \vx_0$ as independent variables and matching the coefficients, we obtain the constraint:
\begin{equation}
    \frac{\sigma_{t-\Delta t}^2}{\sigma_t^2} = \frac{\lambda_t^2}{\tilde{\lambda}_{t-\Delta t}^2}.
\end{equation}

Under the Markov assumption, the conditional distribution $p(\vx_t \mid \vx_{t-\Delta t}, \vx_0)$ must be independent of $\vx_0$. Therefore, from \eqref{eq:ddim-reverse}, the mean term must satisfy:
\begin{equation}
    \mu_t - \tilde{\gamma}_{t-\Delta t} \cdot \frac{\mu_{t-\Delta t}}{\sigma_{t-\Delta t}} = 0,
\end{equation}
which implies:
    $\tilde{\gamma}_{t-\Delta t} = \frac{\mu_t \sigma_{t-\Delta t}}{\mu_{t-\Delta t}}$.

Substituting into the definition of $\tilde{\lambda}_{t-\Delta t}$:
\begin{equation}
    \tilde{\lambda}_{t-\Delta t}^2 
    = \sigma_t^2 - \tilde{\gamma}_{t-\Delta t}^2
    = \sigma_t^2 - \left( \frac{\mu_t \sigma_{t-\Delta t}}{\mu_{t-\Delta t}} \right)^2
    = \sigma_t^2 - \frac{\mu_t^2}{\mu_{t-\Delta t}^2} \sigma_{t-\Delta t}^2.
\end{equation}

Finally, using the earlier ratio between $\lambda_t$ and $\tilde{\lambda}_{t-\Delta t}$, we obtain:
\begin{equation}
    \lambda_t^2 
    = \tilde{\lambda}_{t-\Delta t}^2 \cdot \frac{\sigma_{t-\Delta t}^2}{\sigma_t^2}
    = \frac{\sigma_{t-\Delta t}^2}{\sigma_t^2} \left( \sigma_t^2 - \frac{\mu_t^2}{\mu_{t-\Delta t}^2} \sigma_{t-\Delta t}^2 \right).
\end{equation}
       
   \end{proof} 
 Note that our derivation holds for any noise schedule $\{\mu_t,\sigma_t\}$, without assuming VP or any particular form of $\mu_t$ and $\sigma_t$.  

\begin{proposition}
\label{prop: bfn markov}
We verify that the sampling formulation of BFN for continuous data satisfies the expression for $\lambda_t$ derived under the assumption of a Markov forward process in Proposition \ref{prop: markov}.
\end{proposition}

\begin{proof}
The original BFN sampling procedure for continuous data, as proposed in \cite{graves2023bayesian}, involves multiple iterative parameters. By deriving a closed-form expression, the sampling update can be simplified into the following compact form (see \cite{xue2024unifyingbfn} for detailed derivation):
    \begin{equation}\label{eq:28}
        \vx_{t-\Delta t}=\frac{\mu_{t-\Delta t}}{\mu_t}\vx_t+\frac{\mu_t-\mu_{t-\Delta t}}{\sqrt{\mu_t(1-\mu_t)}}\vepsilon_\theta+\sqrt{\frac{1-\mu_{t-\Delta t}}{1-\mu_t}(\mu_{t-\Delta t}-\mu_t)}\vepsilon.
    \end{equation}
    where $\vepsilon_\theta$ is the predicted noise and $\vepsilon$ is standard Gaussian noise.
    
    To establish its correspondence with the reverse SDE derived under the Markov assumption, it suffices to verify that the mean term in \eqref{eq:28} matches that in \eqref{eq:exact discrete reverse sde}, with the parameter $\gamma_t=\frac{\sigma_{t-\Delta t}^2\mu_t}{\sigma_t\mu_{t-\Delta t}}$ in the Markov forward process case: 
\begin{equation}
\begin{aligned}
    (\frac{\mu_{t-\Delta t}}{\mu_t}\sigma_t-\gamma_t)\sigma_t \vs_\theta=    
    -(\frac{\mu_{t-\Delta t}}{\mu_t}\sigma_t-\frac{\sigma_{t-\Delta t}^2\mu_t}{\sigma_t\mu_{t-\Delta t}}) \vepsilon_\theta
=
    \frac{\mu_t-\mu_{t-\Delta t}}{\sqrt{\mu_t(1-\mu_t)}}\vepsilon_\theta,
\end{aligned}    
\end{equation}
where $\sigma_t^2=\mu_t(1-\mu_t)$ for any $t$ in BFN schedule and $\vs_\theta$ denotes the score function predictor, which is related to the noise predictor via $\vs_\theta(\vx_t, t) = -\vepsilon_\theta(\vx_t, t) / \sigma_t$.
\end{proof}


\end{document}